\documentclass[12pt]{article}

\usepackage{epsfig}

\newcommand{\wt}{\widetilde}

\begin{document}

\title{Theory of doorway states for one-nucleon transfer
reactions. \\
II. Model-independent study of nuclear correlation effects}
\author{B.L. Birbrair \& V.I. Ryazanov\\
PNPI, Gatchina 188350, Russia\\ birbrair@thd.pnpi.spb.ru
 }
\date{} \maketitle

\begin{abstract}
The correlation effects in nuclei owing to which the nuclear
wave functions are different from the Slater determinants are
studied on the basis of the theory of Ref.\cite{1}. The
calculated numbers of nucleons out of the nuclear Fermi-surface
are in reasonable agreement with the finding from the
high-momentum components of the nucleon momentum distributions
in nuclei. The problems concerning the nuclear binding energy
are also discussed.
\end{abstract}

\section{Introduction}

The nucleon-nucleon interaction still remains one of the central
problems of the nuclear structure theory. It seems natural to
use the free-space forces, but the most of the nuclear structure
theorists prefer the effective ones instead. The basic
motivations are as follows.

(a) Historically the first one  arose from the wide-spread
belief in sixties that the free-space $NN$ potential has a hard
repulsive core. Clearly such interaction does not apply directly
because of the divergence of finite-order Feynman diagrams. The
most popular way to overcome this difficulty is to calculate the
Brueckner $G$-matrix and use it as the effective interaction in
the Hartree-Fock problem. This is the Hartree--Fock--Brueckner
approximation, see \cite{2} and the references therein for
details. But soon it became clear that the description of the
two-nucleon system, i.e. the deuteron properties and the elastic
$NN$ scattering phase shifts below the pion production
threshold, does not require the hard repulsive core. As a result
all the contemporary $NN$ potentials are of soft core character,
and so the above difficulty does not really exist. In such
conditions the calculation of the Brueckner $G$-matrix is not
compulsory.

(b) The renormalization of the free-space interaction due to the
medium polarization effects. But such effects are treated by
conventional methods of quantum many-body theory, and therefore
the above reason is not the argument for the effective forces.

(c) The medium QCD-renormalization due to the fact that the
quark composition of the QCD vacuum is changed in nuclear medium
thus leading to the changes of both the mesons and the
meson--nucleon vertices \cite{3}--\cite{6}. Such processes can
hardly be described in all details since the exact theory does
not yet exist for the nonperturbative region. Nevertheless, at
least one exact statement can be done. The QCD-renormalized
interaction is the functional of nuclear density possessing the
following obvious property: it turns into the free-space one in
the zero-density limit. For this reason it can be represented as
the following functional expansion:
\begin{eqnarray}
f_{QCD-r}({\bf r}_1{\bf r}_2\{\rho\}) &=& f_2(|{\bf r}_1-{\bf
r}_2|)+\int f_3(|{\bf r}_1-{\bf r}_2|,|{\bf r}_1-{\bf r}'|)
\rho(r')d^3{\bf r}'\ + \nonumber\\
&+& \int\hspace{-0.4em}\int f_4 (|{\bf r}_1-{\bf r}_2|,|{\bf r}
_1-{\bf r}'|,|{\bf r}_1-{\bf r}''|)\rho(r')\rho(r'')
d^3{\bf r}'d^3{\bf r}''+\ \cdots
\end{eqnarray}
In this way we conclude that the medium QCD-renormalization is
equivalent to the existence of many-particle $NN$ forces in
addition to the two-particle ones. This conclusion is confirmed
by the fact that the physics of strong interaction is
essentially nonlinear. But the nonlinearity automatically leads
to many-particle forces.

The above reasons are grounds for our starting point: both the
two-particle and many-particle $NN$ forces must be taken into
account for the treatment of nuclear structure.

The three-particle $NN$ forces are indeed included for the
calculations  of few-nucleon systems \cite{7}. But the following
questions arises: is this sufficient for complex nuclei~? This
point as well as a number of those concerning the nuclear
structure might be elucidated if the model-independent object
would exist in nuclei.

Our approach is based on the fact that such objects do really
exist. As discussed in our previous work \cite{1} they are the
doorway states for one-nucleon transfer reaction. We showed that

(i) at least the three-particle repulsion and the four-particle
attraction must be taken into account in addition to the
two-particle forces;

(ii) the nuclear relativity is really existing phenomenon rather
than the hypothesis of J.D.~Walecka \cite{8};

(iii) the dominant contribution to the isovector nuclear
potential is provided by the many-particle forces;

(iv) the observed spectra of the doorway states can be used to
specify the neutron density distributions in nuclei; such
specified densities are indeed obtained for the closed-shell
nuclei $^{40}$Ca, $^{90}$Zr and $^{208}$Pb.

In the present work our approach is applied to the "empirical"
studies of nuclear correlation effects owing to which the
nuclear wave functions are different from the Slater
determinants. Such effects are treated by a variety of
approximate methods since the exact ones do not exist. For this
reason it is very important to get a model-independent
quantitative information about the above effects. The
possibility provided by our approach is based on the fact that
the single-particle states of nucleon in nuclear static field
(which are just the doorway states for one-nucleon transfer
reactions \cite{1}) are
correlation-free objects in contrast to the Landau--Migdal
quasiparticles \cite{9} (which include the correlations by
definition) and single-particle states in nuclear Hartree--Fock
calculations with effective forces (where the correlations are
implicitly included in the phenomenological effective force
parameters). So, calculating the correlation-free quantities and
comparing them with the observed ones we get a quantitative
measure of the correlation effects. In Sect.2 this procedure is
applied to the nucleon density distributions whereas the
problems concerning the nuclear binding energy are discussed in
Sect.3.

\section{Density distributions}

The nucleon density distribution in the ground state of nucleus
$A$ is (see \cite{1} for the notations)
\begin{eqnarray}
\rho(r) &=& \langle A_0|\psi^+(x)\psi(x)|A_0\rangle\
=\ \sum^{(A-1)}_j\langle A_0|\psi^+(x)|(A-1)_j\rangle\langle
(A-1)_j|\psi(x)|A_0\rangle\ = \nonumber\\
&=& \sum^{(A-1)}_j \Psi^+_j(x)\Psi_j(x)\ =\ \int_C\
\frac{d\varepsilon}{2\pi i}\ G(x,x;\varepsilon)\ .
\end{eqnarray}
The integration contour $C$ includes the real axis and the
infinite radius semicircle in the upper part of the complex
$\varepsilon$ plane.  As seen from Eq.(1) the density is
expressed through the single-particle amplitudes of the $(A-1)$
nuclear states.  Expanding them over the complete set of the
doorway states,
\begin{equation} \Psi_j(x)\ =\ \sum_\lambda
C^{(\lambda)}_j\psi_\lambda(x) \end{equation} and putting Eq.(3) into
Eq.(2) we get \begin{equation} \rho(r)\ =\ \sum_\lambda\
n_\lambda|\psi_\lambda(x)|^2 +2\sum_{\lambda\nu \atop
\nu>\lambda}\ \rho_{\lambda\nu} \psi^+_\lambda(x)\psi_\nu(x)\ ,
\end{equation}
where
\begin{equation}
n_\lambda=\rho_{\lambda\lambda}=\sum^{(A-1)}_j |C^{(\lambda)}_j
|^2=\sum^{(A-1)}_j s^{(\lambda)}_j\ , \quad \rho_{\lambda\nu}=
\sum^{(A-1)}_j C^{(\lambda)^*}_jC^{(\nu)}_j
\end{equation}
(actually the coefficients $C^{(\lambda)}_j$ are real quantities
if the parity violation due to weak interaction is disregarded).
The diagonal elements $\rho_{\lambda\lambda}=n_\lambda$ are the
doorway state occupation numbers. Indeed, the particle number is
\begin{equation}
N\ =\ \int\rho(r) d^3{\bf r}\ =\ \sum_\lambda n_\lambda\ ,
\end{equation}
since the nondiagonal elements do not contribute because of the
orthogonality between $\psi_\lambda$ and $\psi_\nu$. The
quantities $n_\lambda$ and $\rho_{\lambda\nu}$ obey the
following limitations:
\begin{equation}
0<n_\lambda<1\ , \quad |\rho_{\lambda\nu}|\ <\
\frac12(n_\lambda+n_\nu)\ .
\end{equation}
The first follows from the fact that the doorway states are
distributed over the actual ones of both the $(A-1)$ and $(A+1)$
nuclei, see Eq.(13c) in Ref.\cite{1}, whereas the second is a
consequence of the Cauchy--Bunjakowsky inequality.

The facts $n_\lambda<1$ and $\rho_{\lambda\nu}\neq0$ reflect the
Fermi surface smearing due to the correlation effects. The
quantities $n_\lambda$ and $\rho_{\lambda\nu}$ carrying the
quantitative information about these effects can be found
"empirically" by considering the relation (4) together with the
limitations (7) as equations for $n_\lambda$ and
$\rho_{\lambda\nu}$. The results will be published elsewhere.

\begin{figure}[h]
\centerline{\epsfig{file=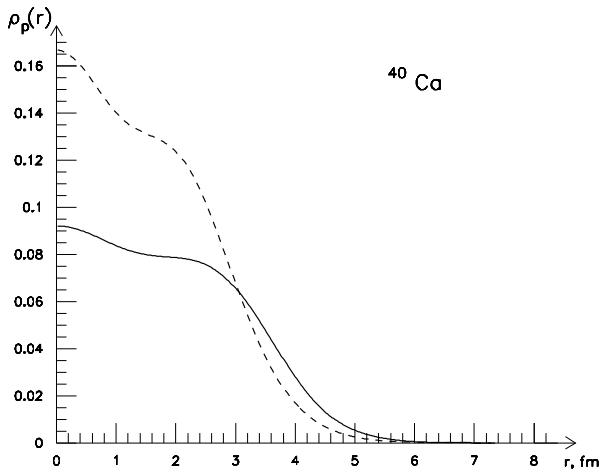,width=8cm}}
\caption{ The observed (full line) and correlation-free
(dashed one) proton density distributions in $^{40}$Ca.}
\end{figure}

As a result of the correlations a part of nucleons is out of the
nuclear Fermi-surface. The number $N_{out}$ of such nucleons is
calculated by comparing the observed density distribution with
the correlation-free one
\begin{equation}
\rho_{cf}(r)\ =\ \sum_\lambda\Theta(\varepsilon_F-\varepsilon_
\lambda)|\psi_\lambda(x)|^2\ .
\end{equation}
$\varepsilon_F$ is the Fermi level energy. Such comparison is
shown in Fig.1 for the proton density distribution in $^{40}$Ca.
As seen from the figure the correlation-free density contains
more nucleons in the inner region $0<r<r_i$ than the observed
one, $r_i$ is the intersection point. The situation in the outer
region $r_i<r<\infty$ is clearly opposite since both densities
correspond to the same number of nucleons. The number of
redistributed nucleons is
\begin{equation}
N_{out}=\ 4\pi\int\limits^{r_i}_0[\rho_{cf}(r)-\rho(r)]r^2dr\ =\
4\pi\int\limits^\infty_{r_i}[\rho(r)-\rho_{cf}(\rho)]r^2dr\ .
\end{equation}
$r_i$ is the intersection point: $\rho_{cf}(r_i)=\rho(r_i)$.
This is just the $N_{out}$ because the only reason for the
redistribution is the Fermi-surface smearing due to the
correlations.

\begin{table}\caption{$N_{out}$ numbers in doubly closed-shell
nuclei}
\begin{center}
\begin{tabular}{ccccc} \hline
& $^{16}$O & $^{40}$Ca & $^{90}$Zr & $^{208}$Pb\\
&&&&\\
$N_{out}$ & 1.10 & 3.15 & 7.34 & 13.15\\

$Z_{out}$ & 1.15 & 3.53 & 6.46 & 13.22\\

$A_{out}$ & 2.25 & 6.68 & 13.80 & 26.37\\

$A_{out}/A$, \% & 14 & 16.7 & 15.3 & 12.7\\
\hline
\end{tabular}
\end{center}
\end{table}

The $N_{out}$ numbers in doubly closed-shell nuclei are shown in
Table 1. As seen from the table these nucleons are a rather
appreciable part of the total mass number. To our knowledge this
fact was first mentioned by Frankfurt and Strikman \cite{10} on
the basis of the analysis of high-momentum components (i.e.
those with $k>300\,$MeV/c) of the nucleon momentum distributions
in nuclei. According to their most recent results for these data
\cite{11} the $A_{out}/A$ ratio is $(20\pm3)\%$ for heavy
nuclei. This is in reasonable agreement with our results.

It is worth mentioning that our calculations do not tell
anything about the nature of the underlying correlations whereas
the high-momentum tales of the momentum distributions arise from
the $NN$ interactions at short distances \cite{10,11}. Therefore
the reasonable agreement between the two results gives rise to
the conclusion that the main reason for the Fermi-surface
smearing in doubly closed-shell nuclei is due to the short-range
correlations.

\section{Binding energy}

We also can calculate the part of the nuclear binding energy
which is caused by the motion of nucleons in nuclear static
field. Such static energy partly includes the correlation
effects since it is expressed through the observed nucleon
density distributions. The comparison between the observed
binding energy ${\cal E}_b$ and the static one ${\cal
E}_{st}$ gives the measure of the proper correlation energy of
nucleus.

To make this point clear let us derive the exact expression for
the binding energy. Following the procedure of Ref.\cite{9} we
get
\begin{eqnarray}
&&  \hspace{-1cm}
{\cal E}_b=\int\!dx\!\int\limits_c\frac{d\varepsilon}{2\pi i}
tr\hat k_xG(x,x;\varepsilon)+\frac12\!
\int\!\!\!\int\!\! dxdx_1\!\int\!\!\!
\int\limits_c\!\frac{d\varepsilon d\varepsilon_1}{(2\pi i)^2}
f_2(|{\bf r}-{\bf r}_1|);\omega)K_2(x,x_1;x,x_1;\varepsilon,
\varepsilon_1)\ + \nonumber\\
+&&\!\!\frac13\int\!\!\!\int\!\!\!\int dxdx_1dx_2
\int\!\!\!\int\limits_c \!\!\!\int
\frac{d\varepsilon d\varepsilon_1d\varepsilon_2}{(2\pi i)^3}f_3
(|{\bf r}-{\bf r}_1|,|{\bf r}-{\bf r}_2|;\omega,\omega_1)K_3
(x,x_1,x_2;x,x_1,x_2;\varepsilon,\varepsilon_1,\varepsilon_2)\
+\nonumber\\
+&& \frac14\!\int\!\!\!\int\!\!\!\int\!\!\!\int\!dxdx_1dx_2dx_3
\!\int\!\!\!\int\!\!\!\int\!\!\!\int\limits_c\!
\frac{d\varepsilon
d\varepsilon_1d\varepsilon_2d\varepsilon_3}{(2\pi i)^4}f_4
(|{\bf r}-{\bf r}_1|,|{\bf r}-{\bf r}_2|,|{\bf r}-{\bf r}_3|;
\omega,\omega_1,\omega_2)\ \times \nonumber\\
\times && K_4(x,x_1,x_2,x_3;x,x_1,x_2,x_3;
\varepsilon,\varepsilon_1,\varepsilon_2,\varepsilon_3)\ ,
\end{eqnarray}
where $K(x_1,\ldots x_n;x'_1\ldots x'_n;\varepsilon_1\ldots
\varepsilon_n)$ are the $n$-particle Green functions. We
accounted for the three- and four-particle forces in addition to
the two-particle ones as well as the possible dependence of the
interactions on the appropriate energy transfers $\omega_i$. In
these terms the single-particle Green function is
\begin{eqnarray}
&& \hspace{-1cm}
(\varepsilon-\hat k_x)G(x,x';\varepsilon)\ =\ \delta(x-x')
+\int dx_1\int\limits_c\frac{d\varepsilon_1}{2\pi i}f_2( |{\bf
r}-{\bf 
r}_1|;\omega)K_2(x,x_1;x',x_1';\varepsilon,\varepsilon_1)\ 
+\nonumber\\
+&& \int\!\!\int dx_1dx_2\int\!\int\limits_c f_3(|{\bf r}-{\bf
r}_1|,|{\bf r}-{\bf r}_2|;\omega,\omega_1)K_3(x,x_1,x_2;x',x_1,
x_2;\varepsilon,\varepsilon_2)\ + \nonumber\\
+&& \int\!\!\int\!\!\int dx_1dx_2dx_3\int\!\!\int\limits_c
\!\!\int
\frac{d\varepsilon_1d\varepsilon_2d\varepsilon_3}{(2\pi i)^3}
f_4(|{\bf r}-{\bf r}_1|,|{\bf r}-{\bf r}_2|,|{\bf r}-{\bf r}_3|;
\omega,\omega_1,\omega_2)\ \times \nonumber\\
\times && K_4(x,x_1,x_2,x_3;x',x_1,x_2,x_3;
\varepsilon,\varepsilon_1,\varepsilon_2,\varepsilon_3)\ .
\end{eqnarray}
On the other hand the Dyson equation is
\begin{equation}
(\varepsilon-\hat k_x)G(x,x';\varepsilon)\ =\ \delta(x-x')+ \int
dx_1M(x,x_1;\varepsilon)G(x_1,x';\varepsilon)\ ,
\end{equation}
and therefore
\begin{eqnarray}
&& \hspace{-1.5cm} \int\!
M(x,x_1;\varepsilon)G(x_1,x';\varepsilon) dx_1=\int dx_1
\int\limits_c\frac{d\varepsilon_1}{2\pi i}f_2(|{\bf r}-{\bf
r}_1);\omega)K_2(x,x_1;x',x_1;\varepsilon,\varepsilon_1)\
+\nonumber\\
+&& \int\!\!\!\int
dx_1dx_2\int\!\!\!\int\limits_c\!\frac{d\varepsilon_1
d\varepsilon_2}{(2\pi i)^2}f_3(|{\bf r}-{\bf r}_1|,|{\bf r}-{\bf
r}_2|;\omega,\omega_1)K_3(x,x_1,x_2;x',x_1,x_2;\varepsilon,
\varepsilon_1,\varepsilon_2)\ + \nonumber\\
+ && \int\!\!\!\int\!\!\!\int dx_1dx_2dx_3\int\!\!\!
\int\limits_c\!\!\!\int\frac{d\varepsilon_1d
\varepsilon_2d\varepsilon_3}{(2\pi
i)^3} f_4(|{\bf r}-{\bf r}_1|,|{\bf r}-{\bf r}_2|,|{\bf r}-{\bf
r}_3|;\omega,\omega_1,\omega_2)\ \times\nonumber\\
\times && K_4(x,x_1,x_2,x_3;x',x_1,x_2,x_3;\varepsilon,
\varepsilon_1,\varepsilon_2,\varepsilon_3)\ .
\end{eqnarray}
As seen from Eqs. (10) and (13) the binding energy can be
written as
\begin{eqnarray}
&& \hspace{-1cm}
{\cal E}_b\ =\ \int\int dxdx'\int\limits_c \frac{d\varepsilon}{
2\pi i}\left(tr\hat k_x\delta(x-x')+M(x,x';\varepsilon)\right)
G(x',x;\varepsilon)\ -\nonumber\\
- && \frac12 \int\int dxdx_1\int\!\!\int\limits_c
\frac{d\varepsilon d
\varepsilon_1}{(2\pi i)^2} f_2(|{\bf r}-{\bf r}_1|;\omega)K_2
(x,x_1;x,x_1;\varepsilon,\varepsilon_1)\ - \nonumber\\
-&& \frac23 \int\!\!\!\int\!\!\!\int\!\!dxdx_1dx_2\!\!
\int\limits_c\!\!\!\int\!\!\!\int
\frac{d\varepsilon d\varepsilon_1d\varepsilon_2}{
(2\pi i)^3}f_3(|{\bf r}-{\bf r}_1|,|{\bf r}-{\bf r}_2|;
\omega,\omega_1) K_3(x,x_1,x_2;x,x_1,x_2;
\varepsilon_1,\varepsilon_2)\ - \nonumber\\
- && \frac34\int\!\!\!\int\!\!\!\int\!\!\!\int dxdx_1dx_2dx_3
\int\!\!\!\int\limits_c\!\!\!\int\!\!\!\int
\frac{d\varepsilon d\varepsilon
_1d\varepsilon_2d\varepsilon_3}{(2\pi i)^4} f_4(|{\bf r}-{\bf
r}_1|,|{\bf r}-{\bf r}_2|,|{\bf r}-{\bf r}_3|,\omega,\omega_1,
\omega_2)\ \times \nonumber\\
\times && K_4(x,x_1,x_2,x_3;x,x_1,x_2,x_3;\varepsilon,
\varepsilon_1,\varepsilon_2,\varepsilon_3)\ .
\end{eqnarray}
Taking into account the spectral representation of $G(x,x';
\varepsilon)$, see Eq.(5) in Ref.\cite{1}, and performing the
integration over $\varepsilon$ we get the following expression
for the first term in the rhs:
\begin{equation}
{\cal E}^{(1)}_b\ =\ \sum^{(A-1)}_j\int dx\Psi^+_j(x)\left( \hat
k_x\Psi_j(x)+\int M(x,x';E_j)\Psi_j(x')dx'\right) .
\end{equation}
As follows from the Dyson equation (12) and the above spectral
representation the amplitudes $\Psi_j(x)$ obey the equation
\begin{equation}
\hat k_x\Psi_j(x)+\int M(x,x';E_j)\Psi_j(x')dx'\ =\ E_j\Psi_j(x)
\end{equation}
and therefore
\begin{equation}
{\cal E}_b^{(1)}=\sum^{(A-1)}_j E_j\int|\Psi_j(x)|^2 dx\ =\
\sum_\lambda\sum^{(A-1)}_j E_js^{(\lambda)}_j
\end{equation}
(we accounted for the expansion (3)).

In general case the many-particle Green functions
$K_n(x_1,\ldots x_n;x'_1,\ldots x'_n;\varepsilon_1,\ldots
\varepsilon_n)$ obey the infinite system of integro-differential
equations the Eq.(11) being the first one. As mentioned above
the exact methods of solution do not exist. It should be also
mentioned that the approximate methods, see \cite{12} and the
references therein, are developed for the instantaneous
two-particle forces only. For these reasons the exact
calculation of nuclear binding energy is impossible at present.
Therefore the calculation of the static energy is of importance.

Both the static nuclear field and the static energy are obtained
from the above relations by putting
\begin{equation}
K_n(x_1,\ldots x_n;x'_1,\ldots x'_n;\varepsilon_1,\ldots
\varepsilon_n)\ =\ \ \prod^n_{i=1} G(x_i,x'_i;\varepsilon_i)\ ,
\end{equation}
i.e. neglecting the difference between the many-particle Green
functions and the unsymmetrized products of the single-particle
ones. This difference arises from both the antisymmetrization
and the higher-order terms of the perturbation theory, i.e. just
the effects leading to the proper correlation energy. Putting
Eq.(18) into Eqs.(13) and (14) we get (all the energy transfers
$\omega_i$ vanish in this case)
\begin{eqnarray}
&& \hspace{-1cm} U_{st}(r)\ =\int f_2(|{\bf r}-{\bf r}_1|)\rho
(r_1)d^3{\bf r}_1+\int\!\!\!\int f_3(|{\bf r}-{\bf r}_1|,|{\bf
r}-{\bf r}_2|)\rho(r_1)\rho(r_2)d^3{\bf r}_1d^3{\bf r}_2\ +
\nonumber\\
+&& \int\!\!\int\!\!\int f_4(|{\bf r}-{\bf r}_1|,|{\bf r}-{\bf
r}_2|,|{\bf r}-{\bf r}_3|)\rho(r_1)\rho(r_2)\rho(r_3)d^3{\bf
r}_1d^3{\bf r}_2d^3{\bf r}_3\ ;\\
&& \hspace{-1cm} {\cal E}_{st}=\sum^{(A-1)}_j\int \Psi^+_j(x)
\left(\hat k_x+U_{st}(r)\right)\Psi_j(x)dx-\frac12\int\!\!\!\int
f_2(|{\bf r}-{\bf r}_1|)\rho(r)\rho(r_1)d^3{\bf r}d^3{\bf r}_1\
- \nonumber\\
- && \frac23\int\!\!\!\int\!\!\!\int f_3(|{\bf r}-{\bf
r}_1|,|{\bf r}-{\bf r}_2|)\rho(r)\rho(r_1)\rho(r_2)d^3{\bf
r}d^3{\bf r}_1 d^3{\bf r}_2\ - \\
- &&
\frac34\int\!\!\!\int\!\!\!\int\!\!\!\int f_4(|{\bf r}-{\bf
r}_1|,|{\bf r}-{\bf r}_2|,|{\bf r}-{\bf r}_3|)\rho(r)\rho(r_1)
\rho(r_2)\rho(r_3)d^3{\bf r}d^3{\bf r}_1d^3{\bf r}_2d^3{\bf
r}_3\ . \nonumber
\end{eqnarray}
As follows from the relations (3) and (5) the first term of the
rhs is
\begin{equation}
{\cal E}_{st}^{(1)}\ =\ \sum_\lambda
n_\lambda\varepsilon_\lambda\ .
\end{equation}
But as seen from Eq.(17) and the sum rule
\begin{equation}
\varepsilon_\lambda\ =\ \sum^{(A-1)}_j E_js^{(\lambda)}_j
+\sum^{(A+1)}_k E_ks^{(\lambda)}_k
\end{equation}
for the doorway state energies, see Eq.(14c) of Ref.\cite{1},
the quantity ${\cal E}^{(1)}_{st}$ contains unphysical
contributions from the energies of the $(A+1)$ nuclear states.
To eliminate them let us divide ${\cal E}^{(1)}_{st}$ into the
two parts
\begin{equation}
{\cal E}^{(1)}_{st}={\cal E}^<_{st}+{\cal E}^>_{st}\ , \quad
{\cal E}^<_{st}=\sum_{\lambda\le F}n_\lambda\varepsilon_\lambda\
, \quad {\cal E}^>_{st}=\sum_{\lambda>F}n_\lambda
\varepsilon_\lambda\ ,
\end{equation}
including summations over the states with $\varepsilon_\lambda
\le\varepsilon_F$ and $\varepsilon_\lambda>\varepsilon_F$
respectively. The doorway states entering ${\cal E}^<_{st}$ are
mainly distributed over the states of the $(A-1)$ nucleus the
first term of Eq.(22) thus giving the dominant contribution to
the $\varepsilon_\lambda$ values in this case. The small
unphysical contribution from the second term of (22) arises from
the states of the $(A+1)$ nucleus with the same quantum numbers
as those of the low-lying states of the $(A-1)$ nucleus. Such
states lie either in the continuum or near its border. Therefore
the reasonable estimate for their energies is $E_k\approx0$ the
unphysical contribution thus being negligible for this case.

The situation is opposite for ${\cal E}^>_{st}$ because the
doorway states with $\varepsilon_\lambda>\varepsilon_F$ are
mainly distributed over the states of the $(A+1)$ nucleus, the
dominant contribution to the $\varepsilon_\lambda$ values thus
being unphysical from the viewpoint of the binding energy. In
such conditions it is reasonable to use a more appropriate
relation for ${\cal E}^>_{st}$,
\begin{equation}
\wt{\cal E}\,^>_{st}\ =\ \sum_{\lambda>F} \sum^{(A-1)}_j
E_js^{(\lambda)}_j\ .
\end{equation}
So the $(A-1)$ nuclear states with the same quantum numbers as
those of the low-lying states of the $(A+1)$ nucleus contribute
in this case. There are no such states among the low-lying ones
of the $(A-1)$ nucleus, and therefore
\begin{equation}
E_j<E_F\ =\ {\cal E}_0(A)-{\cal E}_g(A-1)\ ,
\end{equation}
see Ref.\cite{1} for the details. Nevertheless it is reasonable
to use the estimate
\begin{equation}
E_j\ =\ E_F\ ,
\end{equation}
providing the least absolute value for the $\wt{\cal
E}\,^>_{st}$ quantity. In this way we get
\begin{equation}
\wt{\cal E}\,^>_{st}\ =\ E_F\sum_{\lambda>F}\sum^{(A-1)}_j
s^{(\lambda)}_j\ =\ E_F\sum_{\lambda>F}n_\lambda\ =\ E_F\wt
N_{out}\ .
\end{equation}
Accounting for the fact that the many-particle forces are
introduced as the contact ones in the model-independent approach
of Ref.\cite{1}, we finally get (see Ref.\cite{1} for the
notations)
\begin{eqnarray}
&& \hspace{-1cm} {\cal E}_{st}=\sum^p_{i=n}\left(\sum_{\lambda
\le F} n_\lambda\varepsilon_\lambda+\wt N_{out}E_F\right)_i\ -
\nonumber\\
-&& \frac12\int\left(S(r)\rho_s(r)+V_\omega(r)\rho(r)+S^-(r)
\rho^-_s(r)+V_\rho(r)\wt\rho^-(r)+C(r)\rho_{ch}(r) \right)d^3{\bf
r}\quad - \nonumber \\
- && \int\left(\frac23 a_3\rho^3(r)+\frac34a_4\rho^4(r)+\frac12
\left[a^-_3\rho(r)+a^-_4\rho^2(r)\right](\rho^-(r))^2\right]d^3
{\bf r}\ .
\end{eqnarray}
We used the following ansatz for the occupation numbers
\begin{equation}
n_\lambda\ =\ \frac12\left[1-\frac{\varepsilon_\lambda-\mu}{
\sqrt{(\varepsilon_\lambda-\mu)^2+C^2}}\right], \quad
\sum_\lambda n_\lambda=N, \quad \sum_{\lambda>F}n_\lambda=\wt
N_{out}\ ,
\end{equation}
putting $\wt N_{out}=N_{out}$, see Sect.2. This is incorrect
because the sum $\sum_{\lambda>F}n_\lambda$ is different from
$N_{out}$, Eq.(9). Indeed, both diagonal and nondiagonal
elements of the density matrix, Eq.(4), contribute to the
$N_{out}$ values. This shortcoming will be corrected in future
by calculating the $n_\lambda$ and $\rho_{\lambda\nu}$ values,
see the discussion in Sect.2 (of course, the ansatz (29) will
become unnecessary).

\begin{table}
\caption{The static and observed binding energies in MeV
together with the dominant contributions to ${\cal E}_{st}$.
Those from the Coulomb and the isovector terms are not shown,
but they are included into the ${\cal E}_{st}$ value. The
notations correspond to the terms of Eq.(28).}
\begin{center}
\begin{tabular}{rrrrrrrr} \hline\hline
& ${\cal E}_{st}^{(1)}$\hspace{0.2cm} &
$-\frac12S$\hspace{0.2cm} & $-\frac12V_\omega$\hspace{0.2cm} &
$-\frac23a_3$\hspace{0.2cm} & $-\frac34a_4$\hspace{0.2cm} &
${\cal E}_{st}$ & ${\cal E}_b\; \frac{}{}$\\ \hline
$^{16}$O & $-345.6$ &
+1528.4 & --1224.0 & $-451.2$ & $+461.8$ & $-47.0$ & $-127.6$\\
$^{40}$Ca & $-991.7$ & $+4456.0$ & $-3570.8$ & $-1346.0$ &
$+1411.3$ & $-120.7$ & $-342.0$\\ $^{90}$Zr & $-2251.6$ &
$+10699.0$ & $-8565.6$ & $-3236.5$ & $+3386.0$ & $-234.2$ &
--783.9 \\ $^{208}$Pb & $-4998.2$ & $+26463.4$ & $-21255.6$ &
$-8190.9$ & $+8694.0$ & $-233.3$ & $-1636.5$ \\ \hline\hline
\end{tabular} \end{center} \end{table}

The results of the calculations are shown in Table 2. As seen
from the table the static energy is the sum of a number of
contributions with different sign, the dominant ones greatly
exceeding the  total static energy as well as the observed
binding one. This is the source of the ambiguities because all
disregarded effects may be of importance in such a situation.

One of such effects is the finite range of the many-particle
forces. The possible way of the estimation is illustrated for
the three-particle forces. As seen from Eq.(10) the original
contribution is
\begin{eqnarray}
&& \hspace{-1.5cm} {\cal E}_3\ =\ \frac13\int\!\!\!\int\!\!\!\int
f_3(|{\bf r}-{\bf r}_1|,|{\bf r}-{\bf r}_2|)\rho(r)\rho(r_1)
\rho(r_2)d^3{\bf r}d^3{\bf r}_1d^3{\bf r}_2\ =\nonumber\\
= && \frac13\int\!\!\!\int\!\!\!\int \rho(r)f_3(\xi,\eta)
\rho(|{\bf r}+\mbox{\boldmath$\xi$}|)\rho(|{\bf r}+
\mbox{\boldmath$\eta$}|)d^3{\bf r}d^3\mbox{\boldmath$\xi$}
d^3\mbox{\boldmath$\eta$}\ \cong \nonumber\\
\cong && \frac13\int\!\!\!\int\!\!\!\int\rho(r)f_3(\xi,\eta)
\left[\rho(r)+\mbox{\boldmath$\xi\nabla$}\rho(r) +\frac12
\xi_1\xi_k\partial_i\partial_k\rho(r)+\cdots\right]\
\times \nonumber\\
\times && \left[\rho(r)+\mbox{\boldmath$\eta\nabla$}\rho(r)
+\frac12\eta_j\eta_\ell\partial_j\partial_\ell\rho(r)+\cdots\right]
d^3{\bf r}d^3\mbox{\boldmath$\xi$}d^3\mbox{\boldmath$\eta$}\
=\\
=&&\frac13\int\!\!\!\int\!\!\!\int\rho(r)f_3(\xi,\eta)\left[
\rho(r)+\frac16\xi^2\Delta\rho(r)+\cdots\right]\left[\rho(r)
+\frac16\eta^2\Delta\rho(r)+\cdots\right]d^3{\bf r}d^3
\mbox{\boldmath$\xi$}d^3\mbox{\boldmath$\eta$}\ = \nonumber\\
= && \frac13 a_3\int\rho(r)[\rho(r)+r^2_0\Delta\rho(r)+\cdots]^2
d^3{\bf r}\cong\frac13a_3\int\rho^3(r)d^3{\bf r}+\frac23a_3\int
\rho^2(r)\Delta\rho(r)d^3{\bf r}\ ,\nonumber
\end{eqnarray}
where (see Eq.(37) of Ref.\cite{1})
\begin{equation}
a_3=\int\!\!\!\int f_3(\xi,\eta)d^3\mbox{\boldmath$\xi$}
d^3\mbox{\boldmath$\eta$}\ , \quad a_3r^2_0=\frac16\int\!\!\!\int
\xi^2f_3(\xi,\eta)d^3\mbox{\boldmath$\xi$}
d^3\mbox{\boldmath$\eta$}
=\frac16\int\!\!\!\int\eta^2f_3(\xi,\eta)d^3\mbox{\boldmath$\xi$}
d^3\mbox{\boldmath$\eta$}\ .
\end{equation}
In this way we get
\begin{eqnarray}
&& \delta_{r^2_0}{\cal E}_{st}=r^2_0\int\left(\frac23
a_3\rho^2(r)+\frac34a_4\rho^3(r)\right)\Delta\rho(r)d^3{\bf r}=
-Dr^2_0 \nonumber\\
&& D\ =\ \int\left(\frac43a_3\rho(r)+\frac94a_4\rho^2(r) \right)
(\rho'(r))^2 d^3\bf r\ .
\end{eqnarray}

\begin{table} \caption{The many-particle finite range effect}
 \begin{center}
\begin{tabular}{ccccc} \hline\hline

& $^{16}$O & $^{40}$Ca & $^{90}$Zr & $^{208}$Pb\\
&&&&\\
$D$, MeV$\cdot$fm$^{-2}$ & $-12.12$ & $-3.29$ & 22.02 & 34.08\\
&&&&\\ \hline\hline
\end{tabular}\end{center}\end{table}

The calculated $D$ values are shown in Table 3. The reasonable
value of the range parameter is the $\omega$ meson Compton
wavelength, i.e. a typical scale of the strong interaction. So
$r^2_0=m^{-2}_\omega=0.1\,$fm$^2$, the effect thus being
negligibly small.

The second source of the possible ambiguity is seen from the
results of Ref.\cite{1} for the contributions of two-particle,
three-particle and four-particle forces to the nuclear static
field which are shown in Table 4, see Eqs. (44) and (45) of
Ref.\cite{1}.

\begin{table} \caption{The contributions to the static field}
 \begin{center}\begin{tabular}{lccc} \hline\hline

& $U_2$ & $U_3$ & $U_4$ \\ &&&\\
Bonn B & $-83$ & $+96.5$ & $-104$ \\
OSBEP & $-82$ & $+97$ & $-107$ \\
\hline\hline \end{tabular}\end{center}\end{table}

As follows from the table the convergence of this sequence is
not seen. This observation suggests the possible existence of
the contributions from higher many-particle forces, and
therefore the whole sequence should be summed up. But this can
be done only within some reasonable model for the many-particle
forces including all higher-order ones as well as the finite
range and the mechanism for the saturation of nuclear density.
The corresponding investigation is in progress.

\end{document}